\documentclass[prd,aps,twocolumn,nofootinbib,preprintnumbers,showpacs]{revtex4}
\usepackage{amsmath,amssymb,epsf}
\def\bq{\begin{quote}}
\def\eq{\end{quote}}

\begin{document}
\title{Eternal inflation and localization on the landscape}
\author{D. Podolsky${}^1$\footnote{On leave from Landau Institute
for Theoretical Physics, 119940, Moscow,
Russia.} and K. Enqvist${}^{1,2}$}
\affiliation{${}^1$ Helsinki Institute of Physics, P.O. Box 64
(Gustaf H\"{a}llstr\"{o}min katu 2), FIN-00014, University of
Helsinki, Finland}
\affiliation{${}^2$ Department of Physical Sciences, P.O. Box 64, FIN-00014, University of Helsinki, Finland}
\date{\today}
\begin{abstract}
We model the essential features of eternal inflation on the
landscape of a dense discretuum of vacua by the potential
$V(\phi)=V_{0}+\delta V(\phi)$, where $|\delta V(\phi)|\ll V_{0}$ is
random. We find that the diffusion of the distribution function
$\rho(\phi,t)$ of the inflaton expectation value in different Hubble patches
%is strongly
may be suppressed due to the effect analogous to the Anderson localization in disordered quantum systems.
At $t \to \infty$ only the localized part of the distribution function
$\rho (\phi, t)$ survives which leads to dynamical selection principle on the landscape.
The probability to measure any but a small value of the cosmological constant in a given Hubble patch on the landscape is exponentially suppressed at $t\to \infty$.
\end{abstract}
\maketitle

String theory is believed to imply a wide landscape \cite{susskind} of
both metastable vacua with a positive cosmological constant and true
vacua with a vanishing or a negative cosmological constant; the latter
are called anti-de Sitter or AdS vacua, where space-time collapses
into a singularity.
In regions with positive cosmological constant, or in de Sitter (dS) vacua,
the universe inflates, and because of
the possibility of tunneling between different de Sitter vacua inflation is eternal.

The problem of calculating statistical
distributions of the landscape vacua is very complicated
\cite{DouglasStatistics} and is even considered to be NP-hard
\cite{NPhard} (the total number of vacua on the landscape is estimated to be of order
$10^{100}\div10^{1000}$).
Our aim is to consider how eternal inflation
proceeds on the landscape by using the mere fact that the
number of vacua within the landscape is extremely large, so that
their distribution can have significant disorder. The dynamics
of eternal inflation is then
described by the Fokker-Planck equations in the disordered effective potential.\footnote{An approach somewhat similar to ours was also presented in \cite{EastherRandom}.}
In that case, the landscape dynamics may have some interesting parallels in
solid state physics, as we will discuss
in the present paper.

Eternal inflation on the landscape can be modeled as follows \cite{LindeLandscapeEarly,LindeLandscapeLate}.
Let us numerate vacua on the landscape by the discrete index $i$
and define $P_{i}(t)$ as the probability to measure a given (positive) value of the cosmological constant $\Lambda_{i}$ in a given Hubble patch.
If the rates of tunneling between the metastable minima $i$ and $j$ on the landscape are given by the time independent matrix $\Gamma_{ij}$, then the probabilities $P_{i}$ satisfy the system of ``vacuum dynamics''
equations \cite{VacuumDynEq}
\begin{equation}
\dot{P}_{i}=\sum_{j\ne i}\left(\Gamma_{ji}P_{j}-\Gamma_{ij}P_{i}\right)-\Gamma_{is}P_{i}.
\label{eq:VacDynFull}
\end{equation}
The last term in this equation corresponds to tunneling between the metastable de Sitter
vacuum $i$ and a true vacuum with a negative cosmological constant (an AdS vacuum), i.e. tunneling
into a collapsing AdS space-time \cite{AdSTunnel}. The collapse time $t_{\rm col}\sim {M_P}/{V_{\rm AdS}^{1/2}}$
is much shorter than the characteristic time
$t_{\rm rec}\sim \exp \left( {M_P^4}/{V_{\rm dS}} \right)$ for tunneling back
into a de Sitter metastable vacuum, so that the AdS true vacua effectively
play the role of sinks for the probability current (\ref{eq:VacDynFull}) describing eternal
inflation on the landscape \cite{LindeLandscapeEarly}.

In what follows we will assume that the effect of the AdS
sinks is relatively small; otherwise the landscape will be divided
into almost unconnected ``islands'' of vacua \cite{LindeLandscapeLate}, preventing
the population of the whole landscape by eternal inflation.

In the limit of weak tunneling only the vacua closest to each other are
important. It is convenient
to classify parts (islands) of the landscape according to the typical
number of adjacent vacua within each part. Technically, the landscape of vacua of the string theory can be represented as a graph with $10^{100}\div{}10^{1000}$ nodes and a number of connections
between them of the same order. By an island on the landscape, we
mean a subgraph relatively weakly connected to the major
"tree". The dimensionality of the island can then be defined as the
Hausdorff
dimension $N_H$ of the corresponding subgraph \cite{PodolskyFullForthcoming}.
For example, if there are only two adjacent vacua for any vacuum in a given island, then $N_H=1$
for this island and we denote it as quasi-one-dimensional; a domain of vacua with $N_H=2$ is quasi-two-dimensional,
and so on.

In the quasi-one-dimensional case (neglecting the
AdS sinks) the system (\ref{eq:VacDynFull}) reduces
to
\begin{equation}
\dot{P}_{i}=-\Gamma_{i,i+1}P_{i}+\Gamma_{i+1,i}P_{i+1}-\Gamma_{i,i-1}P_{i}+\Gamma_{i-1,i}P_{i-1}.
\label{eq:DynVacSimple}
\end{equation}
While in general $\Gamma_{ij}\ne\Gamma_{ji}$, we will take
$\langle\Gamma_{ij}\rangle=\langle\Gamma_{ji}\rangle$
on the average.\footnote{This condition is never satisfied for the Bousso-Polchinski landscape \cite{BoussoPolchinski}, where the adjacent vacua are those with closest values of the effective cosmological constant. However, the spectrum of states on Bousso-Polchinski landscape is not disordered, so that the analysis based on averaging over disorder is not applicable.  Disorder appears in more realistic multithroat models of the string theory landscape.} Furthermore, suppose that
the initial condition for Eq. (\ref{eq:DynVacSimple}) is
\begin{equation}
P_{i}(0)=1,\,\, P_{j\ne i}(0)=0.\label{eq:VacDynInitialCond}
\end{equation}
so that the initial state is well localized.
Naively, one may expect
that the distribution function $P_{i}(t)$ would start to spread out
according to the usual diffusion law and the
system of vacua would exponentially quickly reach a {}``thermal'' equilibrium distribution
of probabilities for a given Hubble patch to be in a given dS vacuum.
However, there exists a well known
theorem \cite{Sinai} from the theory of diffusion on random
lattices stating that the distribution
function $P_i$ remains localized near the initial distribution peak for
a very long time, with its characteristic width behaving as
\begin{equation}
\langle i^{2}(t)\rangle\sim\log^{4}t~.
\label{eq:LatticeSpread}
\end{equation}
This is a surprising result when applied to eternal inflation where the
general lore (see for example \cite{InfInitial}) is that the
initial conditions for eternal inflation will be forgotten almost
immediately after its beginning. Instead, in what follows we will argue that
the memory about the initial conditions may survive during a very long time on the quasi-one-dimensional
islands of the landscape.

We will model the landscape by a continuous inflaton potential
\begin{equation}
V(\phi)=V_{0}+\delta V(\phi),
\label{eq:simplepotential}
\end{equation}
where $V_{0}$ is constant, and $\delta V(\phi)$ is a random contribution
such that $|\delta V(\phi)|\ll V_{0}$, and $\phi$ is the inflaton or the order
parameter describing the transitions.
As in stochastic inflation \cite{stochasticinflation}, in different causally connected regions fluctuations
have a randomly distributed amplitude and observers living in different
Hubble patches see different expectation values of the inflaton. When
stochastic fluctuations of the inflaton are large enough,
the expectation value of the inflaton in a given Hubble patch is determined by the Langevin equation \cite{stochasticinflation}
\begin{equation}
\dot{\phi}=-\frac{1}{3H_{0}}\frac{\partial\delta V}{\partial\phi}+f(t),
\label{eq:langevineq}
\end{equation}
where the stochastic force $f(\phi,t)$ is Gaussian with correlation
properties
\begin{equation}
\langle f(t)f(t')\rangle=\frac{H_{0}^{3}}{4\pi^{2}}\delta(t-t').
\label{eq:forcedisp}
\end{equation}
From (\ref{eq:langevineq}) one can derive the Fokker-Planck equation,
which controls the evolution of the probability distribution $\rho(\phi,t)$
describing how the values of $\phi$ are distributed among different
Hubble patches in the multiverse.
One finds \cite{stochasticinflation}
\begin{equation}
\frac{\partial\rho(\phi,t)}
{\partial t}=\frac{H_{0}^{3}}{8\pi^{2}}\frac{\partial^{2}\rho}{\partial\phi^{2}}+\frac{1}{3H_{0}}
\frac{\partial}{\partial\phi}\left(\frac{\partial V}{\partial\phi}\rho\right).
\label{eq:fokkerplanck}
\end{equation}
The general solution to Eq. (\ref{eq:fokkerplanck})
is given by
\begin{equation}
\rho=e^{-\frac{4\pi^{2}\delta V(\phi)}{3H_{0}^{4}}}\sum_{n}c_{n}\psi_{n}(\phi)e^{-\frac{E_{n}H_{0}^{3}(t-t_{0})}{4\pi^{2}}},
\label{eq:fpgenralsolution}
\end{equation}
where $\psi_{n}$ and $E_{n}$ are respectively the eigenfunctions
and the eigenvalues of the effective Hamiltonian
\begin{equation}
\hat{H}=-\frac{1}{2}\frac{\partial^{2}}{\partial\phi^{2}}+W(\phi).
\label{eq:FPhamiltonian}
\end{equation}
Here
\begin{equation}
W(\phi)=\frac{8\pi^{4}}{9H_{0}^{8}}\left(\frac{\partial\delta V}{\partial\phi}\right)^{2}-\frac{2\pi^{2}}{3H_{0}^{4}}\frac{\partial^{2}\delta V}{\partial\phi^{2}}
\label{eq:FPsuperpotential}
\end{equation}
is a functional of the scalar field potential $V(\phi)$. It is often
denoted as the superpotential due to its ``supersymmetric'' form:
the Hamiltonian (\ref{eq:FPhamiltonian}) can be rewritten as
$\hat{H}=\hat{Q}{}^{\dagger}\hat{Q}$, where $\hat{Q}=-\partial/\partial\phi+v'(\phi)$
with $v(\phi)=4\pi^{2}\delta V(\phi)/(3H_{0}^{4}$).

The eigenfunctions of the Hamiltonian (\ref{eq:FPhamiltonian}) satisfy the Schr\"{o}dinger
equation
\begin{equation}
\frac{1}{2}\frac{\partial^{2}\psi_{n}}{\partial\phi^{2}}+(E_{n}-W(\phi))\psi_{n}=0,
\label{eq:FPSchrodinger}
\end{equation}
and its solutions have the following well known features \cite{stochasticinflation}:
\begin{enumerate}
\item The eigenvalues of the Hamiltonian (\ref{eq:FPhamiltonian}) are all positive
definite.
\item The contributions from eigenfunctions of excited states $\psi_{n>0}(\phi)$
to the solution Eq. (\ref{eq:fpgenralsolution}) become exponentially
quickly damped with time. However, if one is interested in what happens
at time scales $\Delta t\lesssim 1/E_{n}$, the first $n$ eigenfunctions
should be taken into account. In particular, if the spectrum of the
Hamiltonian (\ref{eq:FPhamiltonian}) \emph{is very dense}, as in
the case of the string theory landscape, knowing the ground state alone is
not enough for complete understanding dynamics of eternal inflation. %
\end{enumerate}
We now recall that the potential $V(\phi)$
is a random function of the inflaton
field and has extremely large number of minima. This allows us to
draw several conclusions about the form of the eigenfunctions
$\psi_{n}(\phi)$ using the formal analogy between Eq. (\ref{eq:FPSchrodinger})
and the time-independent Schr\"{o}dinger
equation describing the motion of carriers in disordered quantum systems
such as semiconductors with impurities.
The physical quantities in disordered systems can be calculated by averaging
over the random potential of the impurities.\footnote{Observe that the typical number of these impurities
varies between $10^{12}$ to $10^{17}$ per cm$^3$ while the number of vacua on the string theory landscape is $10^{100}\div 10^{1000}$.}

A famous consequence of the random potential generated by impurities
in crystalline materials is the strong suppression of the conductivity, known as Anderson localization \cite{Anderson,Efetov}.
This effect is essential in dimensions lower than $3$ and completely
defines the kinetics of carriers in one-dimensional systems. There, impurities
create a random potential for Bloch waves with the correlation properties
\begin{eqnarray}
\langle u(r)u(r')\rangle & = & \frac{1}{\nu\tau}\delta(r-r'),\,\,\langle u(r)\rangle=0,
\label{eq:ImpuritiesCorrProperties}
\end{eqnarray}
where $\tau$ is the mean free path for electrons and $\nu$ is the density of states per one spin degree
of freedom of the electron gas at the Fermi surface.
As a consequence,
in the one-dimensional case all eigenstates of the electron
hamiltonian become localized with
\begin{equation}
\psi_{n}(r)\sim \exp\left(-\frac{|r-r_{n}|}{L}\right)
\label{eq:LocalizationWaveFunc}
\end{equation}
at $t\to\infty$, where $r_{n}$ are the positions of localization centers, and the localization length $L$
is of the order of the mean free path $l_{\tau}=\langle v\rangle\tau$.
As a result, the probability density $\rho(R,t)$
to find electron at the point $R$ at time $t$ asymptotically
approaches the limit $\rho(R)\sim\exp(-R/L)$ for $R\gg L$, or $\rho(R)\sim{\rm Const}$
for $R\ll L$ at $t\to\infty$. The one-dimensional
Anderson localization takes place for \emph{an arbitrarily
weak disorder} and \emph{arbitrary correlation properties} of the
random potential $u(r)$ \cite{Efetov}.

Also, in a two-dimensional case all the electron eigenstates
in a random potential remain localized. However,
the localization length grows exponentially with energy, the rate
of growth being related to the strength of the disorder.
In three-dimensional case, the localization properties of eigenstates
are defined by the Ioffe-Regel-Mott criterion: if the corresponding
eigenvalue of the Hamiltonian of electrons $E_{n}$ satisfies the
condition $E_{n}<E_{g}$ where $E_{g}$ is so called mobility edge,
then the eigenstate is localized.
The mobility edge $E_{g}$ is a function of the
strength of the disorder. In  higher dimensional cases the situation is unknown.

Let us now return to the discussion of eternal inflation described
by the Fokker-Planck equation (\ref{eq:fokkerplanck}). Since the
localization is the property of the eigenfunctions of the \emph{time-independent}
hamiltonian (\ref{eq:FPhamiltonian}), it is also a natural
consequence of the effective randomness of the potential of the string
theory landscape.\footnote{The Anderson localization on the landscape of string theory was discussed before in \cite{Mersini} in the context of the Wheeler-deWitt equation in the minisuperspace. The possibility to have the Anderson localization on the landscape was also mentioned in \cite{TyeAnderson}.} The diffusion of the probability distribution (\ref{eq:LatticeSpread}) is suppressed due to the localization of
the eigenfunctions $\psi_{n}(\phi)$ contributing to the overall
solution (\ref{eq:fpgenralsolution}). This counteracts the general wisdom that eternal inflation rapidly washes
out any information of the initial conditions.
Indeed,in the quasi-one-dimensional case all the wave functions
$\psi_{n}(\phi)$ are localized, i.e., for a particular realization of disorder they behave as
\begin{equation}
\psi_{n}(\phi)\sim\exp\left(-\frac{|\phi-\phi_{n}|}{L}\right).
\label{eq:localwavefunction}
\end{equation}
where $\phi_{n}$ define the "localization centers" as in the Eq. (\ref{eq:LocalizationWaveFunc}),
and $L$ is the localization length which is of the same order of magnitude as the {}``mean
free path'' related to the strength of the disorder in the superpotential $W(\phi)$.

Let us now discuss how eternal inflation proceeds on islands where
the typical number of adjacent vacua is larger than
two. In the quasi-two-dimensional case the network of vacua within a given
island is described by a composite index $\vec{i}=(i,j)$. The distribution function $\rho$ for finding
a given value of the cosmological constant in a given Hubble patch
is a two-dimensional matrix. Again, all the eigenstates of the corresponding
tunneling hamiltonian $\hat{H}$ are localized. However, since the
localization length grows exponentially with energy, the distribution
function effectively spreads out almost linearly with
\begin{equation}
\langle\vec{i}^{2}(t)\rangle\sim t\left(1+c_{1}\frac{1}{\log^{\alpha}t}+\cdots\right),
\label{eq:2DSpreading}
\end{equation}
where $\alpha>0$ are constants depending on the correlation properties
of the disorder on the landscape \cite{Fisher}. The low energy
eigenstates (namely, the states with $E < E_g$ where $E_g$ is the mobility edge)
are localized with a relatively small localization length.

In the quasi-higher-dimensional cases the distribution function spreads
out according to the linear diffusion law at intermediate times. Again, there exists a mobility edge $E_{g}$ such that the eigenstates of the tunneling Hamiltonian with energies $E<E_{g}$ are localized.
These low energy eigenstates define the asymptotics
of the distribution function $\rho$ at
\begin{equation}
t\gg E_{g}^{-1}.
\label{eq:MobilityEdge}
\end{equation}
The value of the mobility edge $E_{g}$ strongly depends
on the dimensionality of the island and the strength of the disorder,
and the higher is the dimensionality, the lower is the mobility edge \cite{PodolskyFullForthcoming}.

Localization of the low energy eigenstates in two- and higher-dimensional cases introduces an effective dynamical selection principle for different vacua on the landscape (\ref{eq:simplepotential}): in the asymptotic future, not all of them will be populated, but only those near the localization centers $\phi_n$, and the probability to populate other minima will be suppressed exponentially according to the Eq. (\ref{eq:localwavefunction}).

It is interesting to note that in condensed matter systems the localization centers are typically located near the points where the effective potential has its deepest minima \cite{Efetov}. In the case of eternal inflation, it means that the probability to measure any but \emph{very low} value of the cosmological constant in a given Hubble patch will be exponentially suppressed in the asymptotic future \cite{PodolskyFullForthcoming}.

Finally, we discuss the effect of sinks on the dynamics of
tunneling between the vacua. On the string
theory landscape, dS metastable vacua are typically realized by uplifting
stable AdS vacua (as, for example, in the well known KKLT model \cite{KKLT}).
The probability to tunnel from the uplifted dS state $i$ back into
the AdS vacuum is related to the value of gravitino mass $m_{3/2}$
in the dS state \cite{AdSTunnel} and given by
\begin{equation}
t_{{\rm AdS}}\sim\Gamma_{is}^{-1}\sim\exp\left(\frac{{\rm Const.}M_{P}^{2}}{m_{3/2,i}^{2}}\right).
\label{eq:tAdS}
\end{equation}
The gravitino mass after uplifting \cite{AdSUplift}
has the order of magnitude $m_{3/2,i}\sim{|V_{{\rm AdS},i}|^{1/2}}/{M_{P}}$.
Since at long time scales $V_{{\rm AdS},i}$ can also be regarded
as a random quantity, our analysis of the general solution of ``vacuum
dynamics'' equations (\ref{eq:VacDynFull}) does not have to be
modified in any essential way \cite{PodolskyFullForthcoming}.

In addition to AdS sinks, Hubble patches where eternal inflation has ended (stochastic fluctuations of the inflaton expectation value became smaller than the effect of classical force) also effectively play a role of sinks for the probability current described by the Eq. (\ref{eq:fokkerplanck}). In particular, the Hubble patch we live in is one of such sinks. Related to the effect of sinks, there exists a time scale $t_{\rm end}$ for eternal inflation on the landscape (\ref{eq:simplepotential}) such that the unitarity of the evolution of the probability distribution $\rho$ breaks down at $t \gg t_{\rm end}$ \cite{PodolskyFullForthcoming}. Our discussion remains valid if $t \ll t_{\rm end}$. It is unclear whether the probability distribution $\rho$ has achieved the late time asymptotics in the corner of the landscape we live in.

In summary, we have argued that eternal inflation on the landscape
may lead to a strong localization of the inflaton distribution function among different
Hubble patches. This is a consequence of the high density of the
vacua, which effectively implies a random potential for the
order parameter responsible for inflation. We found that the
inflaton motion is analogous to the motion of carriers in disordered quantum systems,
and there exists an analogue of the Anderson localization for eternal inflation on the landscape.
Physically, this means that not all the vacua on the landscape are populated by eternal inflation in the asymptotic future, but only those near the localization centers of the inflaton effective potential. They are located near the deepest minima of the potential, which implies that the probability to measure any but very low value of the cosmological constant in a given Hubble patch is exponentially suppressed at late times.

\subsection*{Acknowledgements}

The authors belong to the Marie Curie Research Training Network HPRN-CT-2006-035863.
D.P. is thankful to I. Burmistrov, N. Jokela, J. Majumder, M. Skvortsov,
K. Turitsyn ad especially to A.A. Starobinsky for the discussions. K.E. is supported
partly by the Ehrnrooth foundation and the Academy of Finland grant
114419.

\end{document}